# A Data Augmentation Pipeline to Generate Synthetic Labeled Datasets of 3D Echocardiography Images using a GAN

**Cristiana Tiago**
GE Vingmed Ultrasound - GE Healthcare
Horten, Norway
cristiana.tiago@gehealthcare.com

**Andrew Gilbert**
GE Vingmed Ultrasound – GE Healthcare
Horten, Norway

**Ahmed S. Beela**
CARIM School for Cardiovascular Diseases
Maastricht University Medical Center
Maastricht, The Netherlands

**Svein Aarne Aase**
GE Vingmed Ultrasound – GE Healthcare
Horten, Norway

**Sten Roar Snare**
GE Vingmed Ultrasound - GE Healthcare
Horten, Norway

**Jurica Sprem**
GE Vingmed Ultrasound - GE Healthcare
Horten, Norway

**Kristin McLeod**
GE Vingmed Ultrasound - GE Healthcare
Horten, Norway

## ABSTRACT

Due to privacy issues and limited amount of publicly available labeled datasets in the domain of medical imaging, we propose an image generation pipeline to synthesize 3D echocardiographic images with corresponding ground truth labels, to alleviate the need for data collection and for laborious and error-prone human labeling of images for subsequent Deep Learning (DL) tasks. The proposed method utilizes detailed anatomical segmentations of the heart as ground truth label sources. This initial dataset is combined with a second dataset made up of real 3D echocardiographic images to train a Generative Adversarial Network (GAN) to synthesize realistic 3D cardiovascular Ultrasound images paired with ground truth labels. To generate the synthetic 3D dataset, the trained GAN uses high resolution anatomical models from Computed Tomography (CT) as input. A qualitative analysis of the synthesized images showed that the main structures of the heart are well delineated and closely follow the labels obtained from the anatomical models. To assess the usability of these synthetic images for DL tasks, segmentation algorithms were trained to delineate the left ventricle, left atrium, and myocardium. A quantitative analysis of the 3D segmentations given by the models trained with the synthetic images indicated the potential use of this GAN approach to generate 3D synthetic data, use the data to train DL models for different clinical tasks, and therefore tackle the problem of scarcity of 3D labeled echocardiography datasets.

# 1    Introduction

Medical imaging plays a crucial role in optimizing treatment pathways. Saving time when it comes to diagnosis and treatment planning enables the clinicians to focus on more complicated cases.

Many modalities are used to image the heart, such as Computed Tomography (CT), Magnetic Resonance (MR), and Ultrasound imaging, enabling several structural and functional parameters related to the organ's performance to be estimated. Such parameters are the basis of clinical guidelines for diagnosis and treatment planning.

Echocardiography is the specific use of ultrasound to image the heart. This imaging modality is widely used given its advantages of being portable, relatively low-cost, and the fact that the use of ionizing radiation is not required.

Deep Learning (DL), and specifically Convolutional Neural Networks (CNNs), have become extensively applied in medical image analysis because they facilitate the automation of many tedious clinical tasks and workflows such as estimation of ejection fraction, for example. These algorithms are capable of approaching human-level performance (Asch et al., 2019), thus potentially saving clinicians' time without decreasing the quality of care for patients. In fact, clinicians agree that using DL algorithms in the clinical workflow also improves patient access to disease diagnoses, increasing the final diagnosis confidence levels (Scheetz et al., 2021). DL models can be developed to perform numerous medical tasks such as image classification, segmentation and even region/structure detection (Aljuaid and Anwar, 2022)

Echocardiography images can be acquired both in 2D and 3D. Time can also be taken into account, generating videos. 3D echocardiography images can be more difficult to assess than 2D images. However, for some specific application cases, 3D image acquisition brings great advantages since it can offer more accurate and reproducible measurements. One such case is ventricle and atrium volumes (Pérez de Isla et al., 2009). Amongst the causes of lack of annotated 3D echocardiography datasets are the higher complexity to acquire 3D echocardiography images and the fact that 3D is still not part of all echocardiography routine exams. Also, even when 3D images are recorded, delineating the structures in them is a challenging, time consuming, and user dependent task. Taken together and adding the fact that privacy regulations to access medical data are becoming stricter, these can explain why there is a lack of publicly available datasets of such type of images. Therefore, having an approach able to address this image scarcity is necessary. This current lack of 3D medical data and the great need of high quality annotated data required by the DL models impacts the development of such algorithms and therefore the scientific and technological development of the 3D medical imaging field. Synthetic generation of labeled 3D echocardiography images is a DL based approach that provides a solution for this problem.

Synthetic data can help in the development of DL models for image analysis (Shin et al., 2018) and accurate labeling of these images. Furthermore, this approach works as a data augmentation strategy by generating additional data. It is known that creating datasets with a combination of real and synthetic images and use them to train algorithms that tackle medical challenges represents a successful solution to the image scarcity (Chen et al., 2021) problem. Such type of synthetic images even increase the heterogeneity present on these datasets, facilitating a more efficient performance of the trained models as they are exposed to a larger variety of images.

Generative Adversarial Networks (GANs) are specific DL architectures that create models capable of generating medical images closely resembling real images acquired from patients. These deep generative models rely on a generator and a discriminator. While the straightforward GAN discriminator distinguishes between real and fake, i.e., generated, images, the generator not only attempts to deceive the discriminator but also tries to minimize the difference between the generated image and the ground truth.

The generated synthetic images can even be associated with labels facilitating the acquisition of large labeled datasets, eliminating the need for manual annotation, and therefore the variabilities associated with the observer (Chuquicusma et al., 2018), which largely influences the final output (Thorstensen et al., 2010). 3D heart models are a great source of anatomical labels since they capture accurate information about the organ's structures (Banerjee et al., 2016). Different types of models can be used for this purpose, such as animated models, biophysical models, or even anatomical models obtained from different imaging modalities (Segars et al., 2010), (Kainz et al., 2019). Recently, CT models were used as label sources to generate 2D echocardiography (Gilbert et al., 2021) and cardiac MR images (Roy et al., 2019), proving the utility of GANs for this task.

Developing a pipeline to generate synthetic data using GANs to create labeled datasets addresses the immense need for the large volume of data that DL algorithms require during training to perform an image analysis task, eliminates the need to acquire the images from subjects, and saves time of experienced professionals when annotating them, as the anatomical labels can be extracted from anatomical models. Usually, when developing such generative models, imaging artifacts are present and visible on the synthetically generated images. This widely common GAN performance drawback is addressed by applying some image post-processing operations (Perperidis, 2016) on the synthetically generated 3D echocardiography images.

In practice synthetic images can be used to train DL models because they represent a good data augmentation strategy (Chai et al., 2021). For instance, 3D medical image segmentation is the most common example of a medical task to which DL can turn out to be a good application. Labeled datasets made of real images combined with synthetic ones, which even include the respective anatomical labels, become the training dataset for 3D DL models, addressing the problem of sparse 3D medical data availability (Lustermans et al., 2022).

## 1.1    State of the Art

DL has become widely used in medical imaging due to its potential in image segmentation, classification, reconstruction, and synthesis across all imaging modalities. Image synthesis has been a research topic for a few decades now, where some of the more conventional approaches use human-defined rules and assumptions like shape priors, for example (Pedrosa et al., 2017). Also, these image synthesis techniques depend on the imaging modality being considered to perform certain tasks. To tackle these shortcomings, CNNs are now becoming a widely used approach for image synthesis across many medical imaging modalities.

Many reasons motivate medical image generation, both 2D and 3D. Generative algorithms can perform domain translation, with a large applicability when converting images from one imaging modality to a different one, as (Uzunova et al., 2020) showed in their work converting 3D MR and CT brain images. GANs can also be used to generate a ground truth for a given input, as these DL models can be trained in a cyclic way, as is the case of the CycleGAN (Zhu et al., 2017), for example. Additionally, generation of synthetic data used for DL algorithms also motivates the application and development of GAN architectures. Several research groups were able to generate medical images using this methodology as a data augmentation tool, even though most of them were developed under a 2D scenario and focused on a few imaging modalities, mainly MRI and CT. These imaging modalities raise less challenges when compared with Ultrasound due to the nature of the physics behind the acquisition process.

Ultrasound images have an inherent and characteristic speckle pattern and their quality is largely influenced by the scanner, the sonographer, and the patient anatomy. When it comes to generating 3D Ultrasound images a few more challenges arise, with the speckle pattern having to be consistent throughout the whole volume being the main one. The anatomical information present in the generated volume also has to hold this consistency feature.

(Huo et al., 2019) trained a 2D GAN model, SynSegNet, on CT images and unpaired MR labels using a CycleGAN. Similarly, (Gilbert et al., 2021) proposed an approach to synthesize labeled 2D echocardiography images, using anatomical models and a CycleGAN as well. The CycleGAN was proposed by (Zhu et al., 2017) and works under an unpaired scenario: the images from one training domain do not have to be related with the images belonging to the other domain. This GAN learns how to map the images from one to another and vice-versa. The paired version of this GAN is called Pix2pix. (Isola et al., 2017) proposed this image synthesis method which generates images from one domain to the other, and vice-versa, however the images belonging to the training domains are paired.

As mentioned, 3D echocardiographic data is sparser, but these images can be generated using GANs, and then used to train new algorithms. Both (Gilbert et al., 2021) and (Amirrajab et al., 2020) investigated the potential use of GAN synthesized datasets to train CNNs to segment different cardiac structures on different imaging modalities, but these methods were limited to 2D.

(Hu et al., 2017) attempted to generate 2D fetal Ultrasound scan images at certain 3D spatial locations. They concluded that common GAN training problems such as mode collapse occur. (Abbasi-Sureshjani et al., 2020) developed a method to generate 3D labeled Cardiac MR images relying on CT anatomical models to obtain labels for the synthesized images, using a SPADE GAN (Park et al., 2019). More recently, (Cirillo et al., 2020) adapted the original Pix2pix model to generate 3D brain tumor segmentations.

When dealing with medical images, U-Net (Ronneberger et al., 2015) is a widely used CNN model to perform image segmentation, for example, since it provides accurate delineation of several structures on these images. More recently, (Isensee et al., 2021) proposed nnU-Net ("no new net"), which automatically adapts to any new datasets and enables accurate segmentations. nnU-Net can be trained on a 3D scenario and optimizes its performance to new unseen datasets and different segmentation tasks, requiring no human intervention.

Existing work to address the challenges of automatic image recognition, segmentation, and tracking in echocardiography has been mostly focused on 2D imaging. In particular, recent work indicates the potential for applying DL approaches to accurately perform measurements in echocardiography images. (Alsharqi et al., 2018) and (Østvik et al., 2018) used a DL algorithm to segment the myocardium in 2D echocardiographic images, from which the regional motion, and from this the strain, were measured. They showed that motion estimation using CNNs is applicable to echocardiography, even when the networks are trained with synthetic data. This work supports the hypothesis that similar approaches could also work for 3D synthetic data.

A large amount of work has been carried out on medical imaging generation and it still represents a challenge for the research community. To the best of our knowledge, the challenge of synthesizing 3D echocardiography images using GANs did not produce any reproducible results, therefore we propose a framework able to address this need.

## 1.2 Contributions

We propose an approach for synthesizing 3D echocardiography images paired with corresponding anatomical labels suitable as input for training DL image analysis tasks. Thus, the main contributions of the proposed pipeline beyond the state of the art include:

1. The extension of (Gilbert et al., 2021) work from 2D to 3D, adapting it from an unpaired to a paired framework (3D Pix2pix) and proposing an automatic pipeline to generate any number of 3D echocardiography images, tackling the lack of public 3D echocardiography datasets and corresponding labels.
2. The creation of a blueprint of heart models and post-processing methods for optimal generation of 3D synthetic data, creating a generic data augmentation tool, this way addressing the lack of 3D data generation works in echocardiography, since it significantly varies from 2D.
3. The demonstration of the usability of these synthetic datasets for training segmentation models that achieve high performance when applied to real images.

## 2 Methodology

The proposed pipeline is summarized in Fig. 1 and described in the following sections. Section II-A describes the preprocessing stage of annotation of the GAN training images to create anatomical labels for these. The training and inference stages are addressed in Section II-B describing how the GAN model was trained and used to synthesize 3D echocardiography images from CT-based anatomical labels and how different post-processing approaches, as described in Section II-C, were applied to these synthetic images. Next, on Section II-D, details regarding the creation of several synthetic datasets used to train 3D segmentation models are given, followed by Section II-E where the influence of adding real images to the synthetic datasets to train segmentation models is assessed.

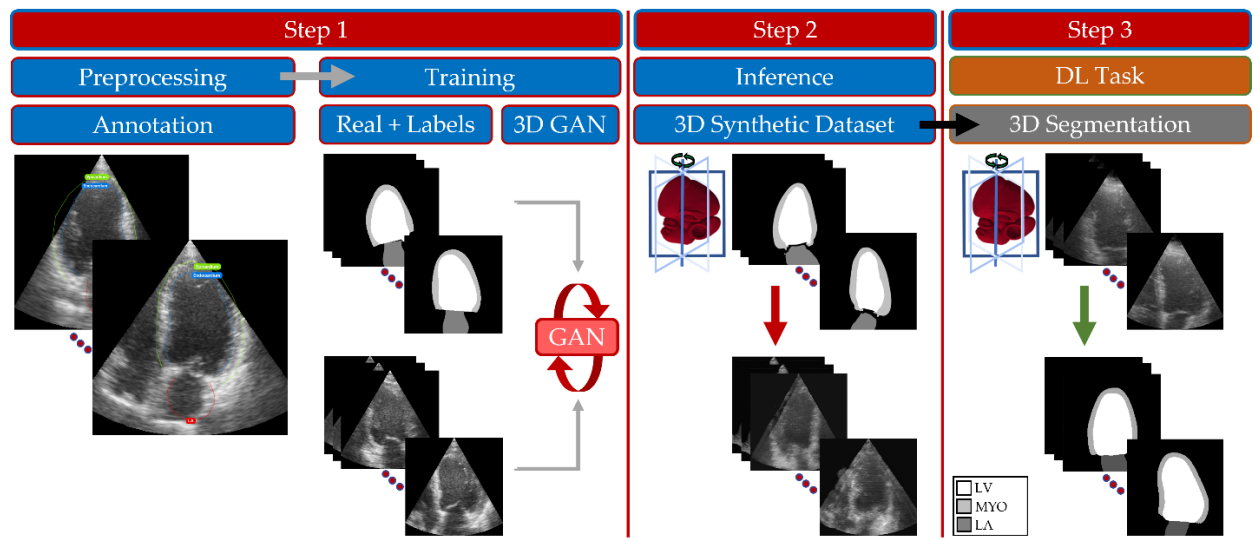

Figure 1: 3D echocardiography image generation pipeline and inference results. Step 1: during the preprocessing stage, a set of 15 3D heart volumes were labeled by a cardiologist and anatomical labels for the LV, LA and MYO were generated. To train the 3D Pix2pix GAN model, the anatomical labels are paired together with the corresponding real 3D images. Step 2: at inference time, the GAN model generates one 3D image. An example obtained during this stage is shown. The proposed method is able to generate physiologically realistic images, giving correct structural features and image details. Step 3: to show the utility of the synthetic datasets, 3D segmentation models were trained using these GAN generated images (black arrow), but other DL tasks can be addressed.

## 2.1 Data Collection

To train the 3D image synthetization model, an annotated dataset was needed since this GAN set up works under a supervised scenario where two sets of images are used for training: a set containing real 3D echocardiography images and a second set with the correspondent anatomical labels manually performed by a cardiologist (see Fig. 1, training stage).

To create the dataset of real 3D echocardiography images, these were acquired during one time point of the cardiac cycle of normal subjects, end-diastole in this work, when the left ventricle (LV) volume is at its largest value. Using GE Vivid Ultrasound scanners 15 heart volumes were acquired.

The second set of images was made up of the anatomical labels corresponding to each of the 3D real images included in the set previously described. Each anatomical label image contains the label for the LV, left atrium (LA), and the myocardium (MYO).

To annotate the 3D echocardiography images a certified member of the American National Board of Echocardiography cardiologist, with more than 10 years of experience, used the V7 annotation tool ("V7 | The AI Data Engine for Computer Vision & Generative AI," n.d.) and contoured the three aforementioned structures (Fig. 1, preprocessing stage) on each of the volumes. These contours were then postprocessed, applying a spline function to the contour points and resampling it, in order to generate gray scale labeled images. All the 3D images present on each training dataset were sized to 256 x 256 x 32.

## 2.2   3D GAN Training

The Pix2pix model was proposed by (Isola et al., 2017) as a solution to image-to-image translation across different imaging domains. This model is capable of generating an output image for each input image by learning a cyclic mapping function across both training domains. The Pix2pix model works as a conditional paired GAN: given two training domains containing paired images, it learns how to generate new instances of each domain. The loss function was kept the same as presented in the original work – a combination of conditional GAN loss and the L1 distance. This way it is conditioning the GAN performance, assuring the information on the generated output image matches the information provided by the input.

This original work was constructed under a 2D scenario, but in this proposed work an extension to 3D was performed by changing the original architecture of the Pix2pix model.

We considered different architectures for the GAN generator and a 3D U-Net (Çiçek et al., 2016) was used to create a 3D version of the Pix2pix model. The discriminator architecture was kept the same, replacing only 2D layers with the correspondent 3D ones. During training of the GAN, data augmentation operations, including blurring and rotation, were performed on the fly, increasing the amount of 3D volumes used without the memory burden of having to save these. The 3D Pix2pix model used here was built using PyTorch (Paszke et al., n.d.) and its training was performed over 200 epochs accounting for the images size and computational memory constraints, considering an initial learning rate of 0.0002 and the Adam optimizer.

At inference time, a common problem among image synthesis is the presence of checkerboard artifacts on the generated images. To tackle this problem, which decreases the quality of the synthesized images, we changed the generator architecture as suggested in (Odena et al., 2016) by replacing the transposed convolutions in the upsampling layers of the 3D U-Net with linear upsampling ones.

In order to generate synthetic echocardiography images for each of the inference cases, i.e., 3D CT-based heart models, the generator part of the GAN, which translates images from the anatomical labels domain to the echocardiography looking images domain, was used. Anatomical models of the heart (Rodero et al., 2021) obtained from CT were used to create the inference gray scale labeled images, containing anatomical information about the LV, LA, and MYO. The main objective of this work was then accomplished by using the GAN as a data augmentation tool to generate synthetic datasets of 3D echocardiography images of size 256 x 256 x 32 from these inference images, augmenting the quantity of 3D echocardiographic image data.

## 2.3   Synthetic Data Post-processing

Tiago *et al.*, A Data Augmentation Pipeline to Generate Synthetic Labeled Datasets of 3D Echocardiography Images using a GANDuring the post-processing stage of the synthetic images generated by the GAN, two different algorithms were experimented. The synthesized images were (a) filtered using the discrete wavelet transform, following (Yadav et al., 2015) work and (b) masked with an Ultrasound cone. The wavelet denoising operation uses wavelets that localize features in the data, preserving important image features while removing unwanted noise, such as checkerboard artifacts. An image mask representing the Ultrasound cone shape was applied to all synthesized images in order to match true Ultrasound data.

## 2.4 3D Segmentation

The GAN pipeline was able to generate labeled instances of 3D echocardiography images, as the model is capable of performing paired domain translation operations. To investigate the utility of the synthetic images, four 3D segmentation models were trained using the generated synthetic images as training set.

The trained model architecture for the 3D segmentation task was the 3D nnU-Net (Isensee et al., 2021). This network architecture was proposed as a self-adapting framework for medical image segmentation. This DL model adapts its training scheme, such as the loss function or slight variations on the model architecture, to the dataset being used and to the segmentation task being performed. It automates necessary adaptations to the dataset such as preprocessing, patch and batch size, and inference settings without the need of user intervention.

To train the first of four 3D segmentation models, $M_{Synthetic}$, described in this section, a labeled dataset made of 27 synthetically generated 3D echocardiography images (256 x 256 x 32), $D_{Synthetic}$, was used. This dataset was obtained from the proposed 3D GAN pipeline at inference time, using anatomical labels from 27 CT 3D anatomical models.

To evaluate the effect of the post-processing operations on the synthesized images, three other datasets were created – $D_{Wavelet}$, $D_{Cone}$, and $D_{WaveletCone}$ – and three additional segmentation models were trained using these – $M_{Wavelet}$, $M_{Cone}$, and $M_{WaveletCone}$, respectively (Fig. 2). $D_{Wavelet}$ was made of the original synthetic images from the $D_{Synthetic}$ dataset but where the wavelet denoising post-processing algorithm was applied, and $D_{Cone}$, was composed by the original synthetic images with the cone reshaping post-processing operation. Finally, a fourth dataset where both post-processing transformations – wavelet denoising and cone reshaping – were applied to the original synthetic images, $D_{WaveletCone}$, was created. All four datasets contained 27 3D echocardiography images with corresponding anatomical labels for the LV, LA and MYO.

All four 3D segmentation models, $M_{Synthetic}$, $M_{Wavelet}$, $M_{Cone}$, and $M_{WaveletCone}$, using nnU-Net, were trained on a 5-fold cross validation scenario during 800 epochs. The initial learning rate was 0.01 and the segmentation models were also built using PyTorch (Paszke et al., n.d.). The loss function was a combination of dice and cross-entropy losses, as described in the original work by (Isensee et al., 2021).

Dice scores were used to assess the quality of the segmentations. This score measures the overlap between the predicted segmentation and the ground truth label extracted from the CT anatomical models. For each segmented structure the Dice score obtained at validation time is a value between 0 and 1, where the latter represents a perfect overlap between the prediction and the ground truth.

## 2.5 Real Data Combined with Synthetic – Data Augmentation

In their work, (Lustermans et al., 2022) showed that adding real data to GAN-generated synthetic datasets can help improve DL models train.

To facilitate a clearer analysis of the influence of using synthetic data to train DL models and the utility of this GAN as a data augmentation tool, three other segmentation models were trained on the datasets $D_{Real}$, $D_{17Real10Augmented}$, and $D_{17Real20Augmented}$. $D_{Real}$ contained 17 real 3D echocardiography volumes acquired with GE Vivid Ultrasound scanners and labeled by a cardiologist.

$D_{17Real10Augmented}$ and $D_{17Real20Augmented}$ were made up of the same 17 real volumes just described together with 10 and 20 synthetic GAN-generated 3D echocardiography images, respectively. Thus allowing to assess the influence of using such type of images during DL models training (Fig. 2).

The 3D segmentation models trained on these datasets were $M_{Real}$, $M_{17Real10Augmented}$, and $M_{17Real20Augmented}$, respectively. All models used the nnU-Net architecture implemented with Pytorch. Similar to the ones described on Section II-D, they were trained for 800 epochs on a 5-fold cross validation scenario, with the same learning rate and loss function.

At inference time, a test set including real 3D echocardiography images was segmented by the three aforementioned models. To compare the segmentation results with the ones obtained from a cardiologist, Dice scores and Volume Similarity (VS) were calculated and used as comparison metrics. VS is calculated as the size of the segmented structures and is of high relevance in a 3D scenario since Dice score presents some limitations. Similarly to the Dice score, this evaluation metric takes values between 0 and 1 but is not overlap-based. Instead, it is a volume based parameter where the absolute volume of a region in one segmentation is compared with the corresponding region volume in the other segmentation (Taha and Hanbury, 2015).

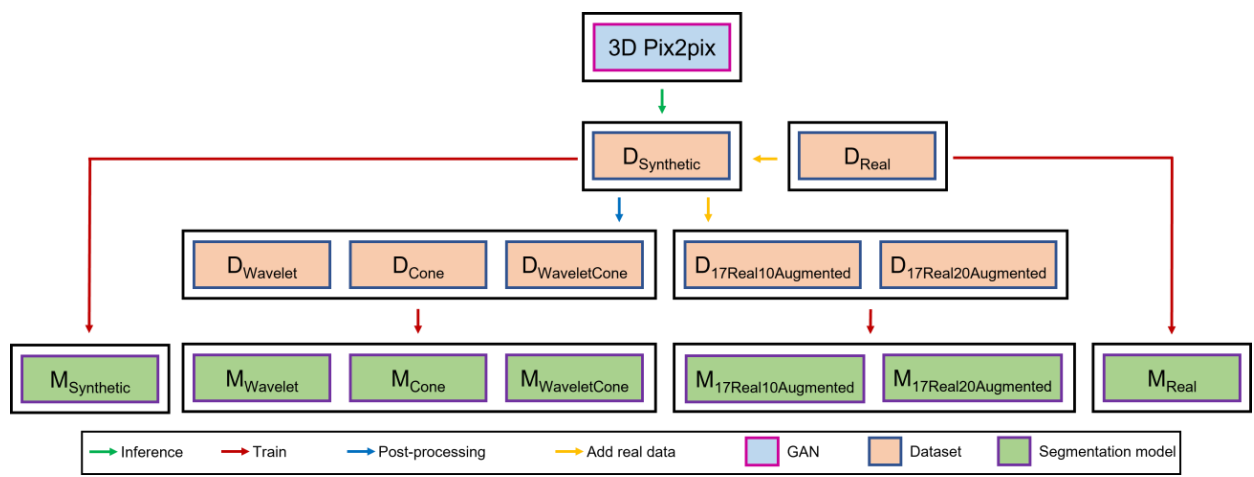

Figure 2: Overview of all the created datasets and trained models in this work. The generative model, 3D Pix2pix, was trained in order to be used to generate synthetic 3D echocardiography datasets. This dataset, $D_{Synthetic}$, was postprocessed applying different transformations and 3 other datasets were created – $D_{Wavelet}$, $D_{Cone}$, and $D_{WaveletCone}$. A fifth dataset completely made of real images, $D_{Real}$, was created and to it, synthetic images from $D_{Synthetic}$ were added creating $D_{17Real10Augmented}$ and $D_{17Real20Augmented}$. All these 7 datasets were used to train 7 3D segmentation models – $M_{Synthetic}$, $M_{Wavelet}$, $M_{Cone}$, $M_{WaveletCone}$, $M_{Real}$, $M_{17Real10Augmented}$, and $M_{17Real20Augmented}$.

## 3 Results

This work's results are presented as follows: Section III-A focuses on the GAN training, architectural modifications performed on the 3D Pix2pix model and their influence on the synthesized images. In Section III-B the influence of post-processing the synthetic images is shown. Finally, Sections III-C and III-D show the segmentation predictions from several models trained on different 3D echocardiography datasets (Fig. 2), as described in Sections II-C and II-D.

### 3.1 GAN Architecture and Training

The chosen GAN architecture influenced the final results. 3D U-Net was chosen as the generator architecture due to its good performance in the medical image domain. The model was trained on a NVIDIA GeForce RTX 2080 Ti GPU and training took five days.

After applying the architectural changes described in Section II-B to remove the checkerboard artifacts, it seemed like these became less visible or even disappeared. However, this correction created some unwanted blurring on the generated images (Fig. 3), therefore the deconvolution layers were used instead of upsampling, and the synthesized images were post-processed to remove the checkerboard artifacts.

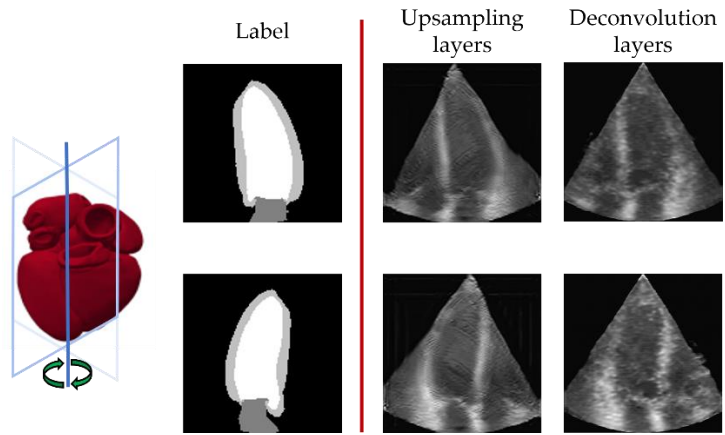

Figure 3: Influence of architectural changes on the GAN generator to remove checkerboard artifacts. At inference time, a 3D anatomical model was used to extract the anatomical labels. The first column shows 2 different slices of this volume at different rotation angles. The middle column shows that synthesizing images using a GAN with upsampling layers smoothens the checkerboard artifacts but introduces blurring, which is not visible on the images when using a GAN with deconvolution layers (right column). Deconvolution layers are preferred to upsampling ones.

### 3.2 Synthetic Data Post-processing

After training the 3D GAN model and generating synthetic images corresponding to the input anatomical models, as described in Section II-C, the obtained 3D echocardiography images were post-processed in order to remove the aforementioned checkerboard artifacts.

The cone edges were slightly wavy in some cases and checkerboard artifacts were sometimes present. The postprocessing experiment, where different transformations were applied to the synthesized images, showed that applying these can give a more realistic aspect to the GAN-generated images, ensuring that the anatomical information remained intact.

Performing these operations allowed to give a more realistic look to the generated echocardiography images (Fig. 4).

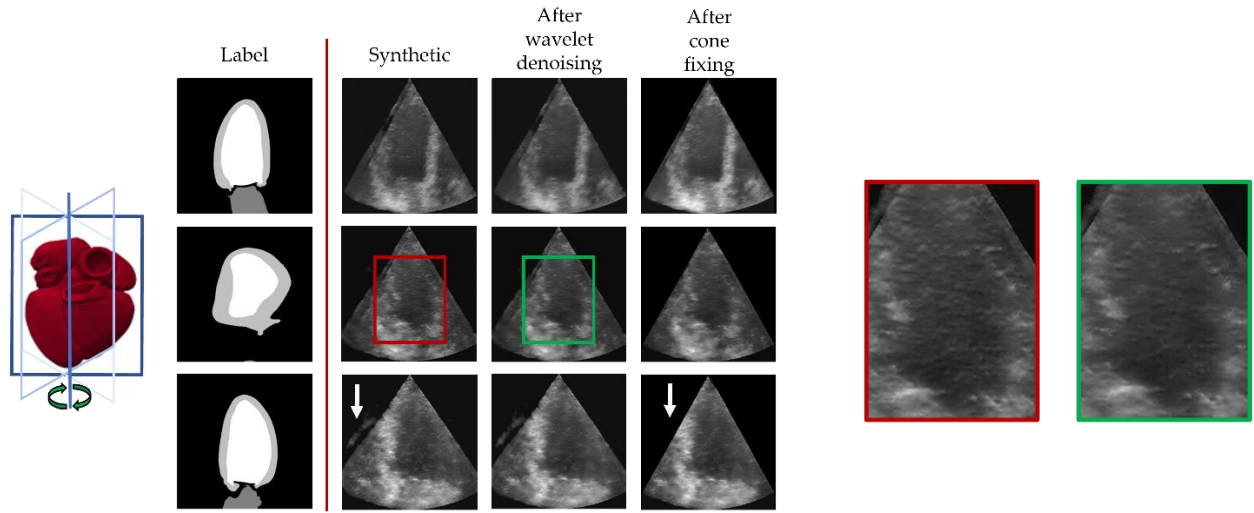

Figure 4: 3D Pix2pix model inference results and post-processing step. At inference time, the anatomical labels were extracted from a 3D heart model. The first column shows 3 different rotation planes of this volume at different rotation angles. After generating the correspondent synthetic ultrasound image (second column) for this inference case, it was post-processed applying a wavelet denoising transformation to eliminate the checkerboard artifacts (third column) and also a cone reshaping step to smooth the wavy edges of the ultrasound cone (fourth column). Post-processing operations give a more realistic look to the synthesized images as indicated by the enlarged areas framed in red and green (wavelet denoise) and the white arrows (cone reshape).

### 3.3 Segmentation from Synthetic Datasets

Anatomical models were used in order to synthesize 27 3D echocardiography images. These were then used to create the synthetic datasets that were used to train 3D segmentation algorithms, as described in section II-D. Post-processing operations were performed on these images to create the $D_{Wavelet}$, $D_{Cone}$, and $D_{WaveletCone}$ datasets. Table 1 shows the average Dice scores (average ± standard deviation) of each segmented structure (LV, LA, and MYO) for each trained model – $M_{Synthetic}$, $M_{Wavelet}$, $M_{Cone}$, and $M_{WaveletCone}$, obtained from the validation dataset. Training took around five days for each fold using a NVIDIA GeForce RTX 2080 GPU, for all epochs. The complete table with all the Dice scores obtained for each training fold of each model can be found in Appendix A – Table 5.

Adding to the Dice scores and to sustain the usability of synthetic images to train segmentation algorithms, Fig. 5 shows the 3D segmentation for an inference 3D echocardiography image acquired from a real subject. Each trained segmentation model was tested on real cases, at inference time.

Table 1: Average validation dice scores (average ± standard deviation) of each segmented structure (LV, LA, and MYO) for each trained model on completely synthetic datasets – $M_{Synthetic}$, $M_{Wavelet}$, $M_{Cone}$, and $M_{WaveletCone}$. The best scores are highlighted.

|  | Models | | | |
| --- | --- | --- | --- | --- |
|  | $M_{Synthetic}$ | $M_{Wavelet}$ | $M_{Cone}$ | $M_{WaveletCone}$ |
| LV | $0.926 \pm 0.006$ | **$0.927 \pm 0.005$** | $0.926 \pm 0.006$ | $0.924 \pm 0.008$ |
| LA | **$0.818 \pm 0.011$** | $0.816 \pm 0.010$ | $0.816 \pm 0.021$ | $0.814 \pm 0.016$ |
| MYO | **$0.808 \pm 0.016$** | $0.808 \pm 0.017$ | $0.803 \pm 0.018$ | $0.801 \pm 0.023$ |

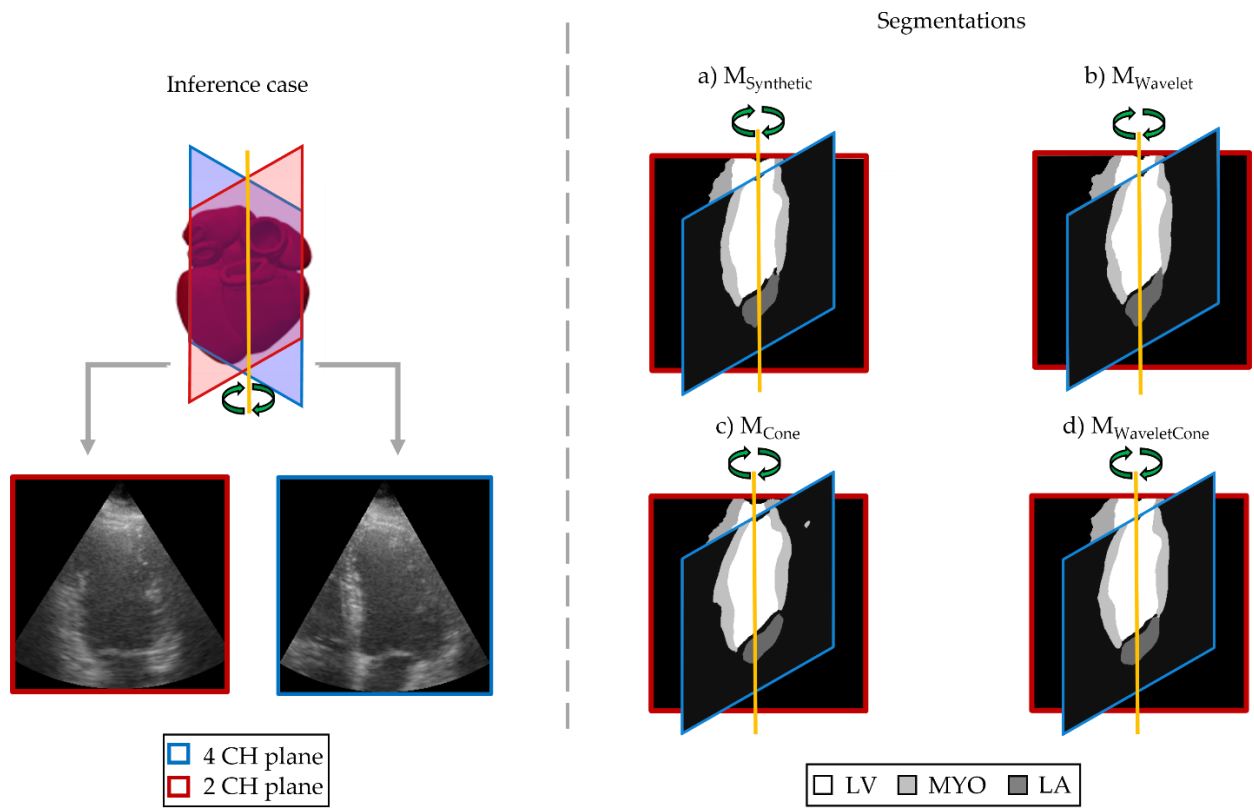

Figure 5: Inference segmentation results from each trained model on synthetic datasets. On the left is shown a schematic representation of the heart and 2 cutting planes correspondent to a real 3D echocardiography image from the test set: the 4-chamber (CH), with blue frame, and the 2-CH, with red frame. On the right, the LV, LA, and MYO segmentation results provided by each of the 4 segmentation models: a) $M_{Synthetic}$, b) $M_{Wavelet}$, c) $M_{Cone}$, and d) $M_{WaveletCone}$ follow. A qualitative analysis of the segmentation results from each of the models, shows that the one where the training data was not post-processed, $M_{Synthetic}$, gives the best output due to a smoother segmentation of the relevant structures.

### 3.4 Segmentation from Combined Datasets

In Table 2 one can see the average Dice scores (average ± standard deviation), obtained at validation time, of each segmented structure (LV, LA, and MYO) for each trained model on the combined datasets: $M_{Real}$, $M_{17Real10Augmented}$, and $M_{17Real20Augmented}$. In Appendix A – Table 4 the complete table with all the Dice scores for each trained fold of all three models can be found.

Table 2: Average validation dice scores (average ± standard deviation) of each segmented structure (LV, LA, and MYO) for each trained model on combined datasets – $M_{Real}$, $M_{17Real10Augmented}$, and $M_{17Real20Augmented}$. The best scores are highlighted.

|  | Models | | |
| --- | --- | --- | --- |
|  | $M_{Real}$ | $M_{17Real10Augmented}$ | $M_{17Real20Augmented}$ |
| LV | **0.938 ± 0.008** | 0.928 ± 0.006 | 0.927 ± 0.007 |
| LA | **0.862 ± 0.023** | 0.830 ± 0.016 | 0.826 ± 0.017 |
| MYO | 0.724 ± 0.028 | **0.767 ± 0.027** | 0.763 ± 0.025 |

Fig. 6 shows the predicted segmentations given by these trained models, next to the ground truth segmentation provided by a cardiologist. The models were tested on a test set made of 3D echocardiography images from real subjects. To compare the output segmentation from the DL models, the Dice scores and VS were calculated based on the predicted segmentations and the anatomical labels from a cardiologist and the results are in Table 3.

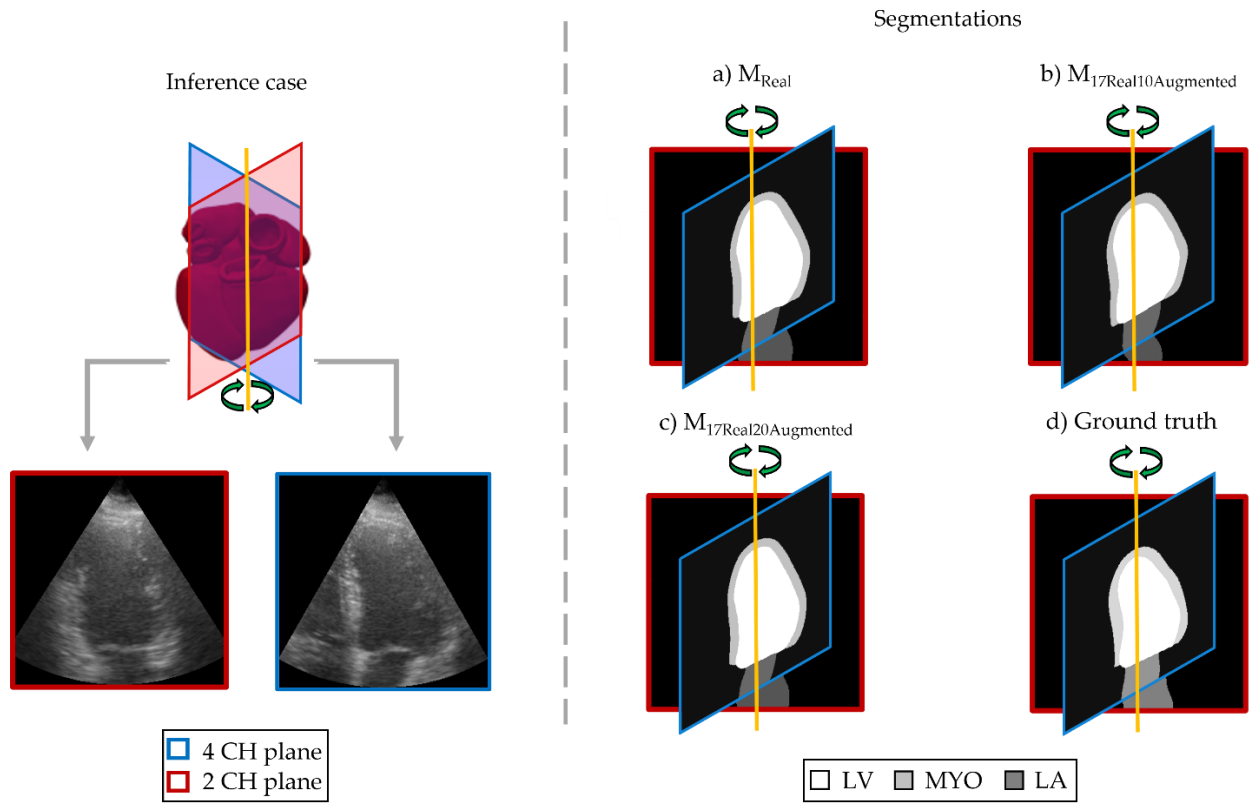

Figure 6: Inference segmentation results from the trained models on augmented datasets with synthetic images. On the left is shown a schematic representation of the heart and 2 cutting planes correspondent to a real 3D echocardiography image from the test set: the 4-CH, with blue frame, and the 2-CH, with red frame. On the right, the LV, LA, and MYO segmentation results provided by the following 3 segmentation models: a) $M_{Real}$, b) $M_{17Real10Augmented}$, and c) $M_{17Real20Augmented}$, follow. To allow comparison and measure the Dice score and VS, d) shows the ground truth segmentation performed by a cardiologist. A qualitative analysis of the segmentation results from each of the models, shows that combining synthetic with real data improves the segmentation output due to a more accurate segmentation of the relevant structures.

Table 3: Average test set dice scores (average ± standard deviation) of each segmented structure (LV, LA, and MYO) and Volume Similarity of the segmented volume for the $M_{Real}$, $M_{17Real10Augmented}$, and $M_{17Real20Augmented}$ models. The best scores are highlighted.

|  | Models | | |
| --- | --- | --- | --- |
|  | $M_{Real}$ | $M_{17Real10Augmented}$ | $M_{17Real20Augmented}$ |
|  | Dice score | | |
| LV | 0.924 ± 0.019 | **0.929 ± 0.020** | 0.922 ± 0.017 |
| LA | **0.876 ± 0.023** | 0.874 ± 0.020 | 0.867 ± 0.022 |
| MYO | 0.666 ± 0.041 | **0.708 ± 0.053** | 0.680 ± 0.063 |
|  | Volume Similarity | | |
| Heart Volume | 0.831 ± 0.038 | **0.844 ± 0.047** | 0.836 ± 0.041 |

# 4 Discussion

In this work we built a pipeline to generate synthetic 3D labeled echocardiography images using a GAN model. These realistic-looking synthetic datasets were used to train 3D DL models to segment the LV, LA, and MYO.

Moreover, combined datasets including synthetic and real 3D images were created, with the VS metric supporting that generated 3D echocardiography images can be used to train DL models, as data augmentation. Segmentation tasks were considered to exemplify the utility of the synthesized data, however the pipeline is generic and could be applied to generate other imaging data and train any DL tasks with anatomical labels as input, as further discussed in this section.

A brief discussion on future applications and modifications of this approach is also presented.

## 4.1 3D Pix2pix GAN - Qualitative Analysis

The pipeline synthesizes 3D echocardiographic datasets with corresponding labels delineating different structures in the images.

After training the 3D Pix2pix GAN model, a qualitative analysis of the synthesized images indicated that the main structures of the heart were well delineated in the generated images (Fig. 1, inference stage). Moreover, image details such as the cone, noise, and speckle patterns are also present and are continuous throughout each volume.

## 4.2 Post-processing and 3D Segmentation - Synthetic Datasets

To evaluate the utilization of synthetic images for research purposes and the extent to which the post-processing transformations affected the final results, four segmentation models were trained using four different datasets, as described earlier in Section III-C.

Despite the very small differences in the Dice scores shown in Table 1 and in Appendix A – Table 5, the inference segmentations (Fig. 5) support the idea that the model trained on the dataset whose images were not post-processed, $M_{Synthetic}$, provided the best segmentation prediction.

The results regarding the influence of the post-processing step on the synthetically generated images supported the fact that applying a wavelet denoising transformation or cone reshaping, or even both transformations together, to these, in order to make the synthetic images look even more realistic, does not necessarily lead to better results when segmenting the LV, LA, and MYO (Fig. 5). This result shows some dependence on the DL task being performed. We segmented large volumes of the 3D image, comparing to its whole content. For this reason, the subtle differences in the voxels intensities that create the checkerboard artifacts do not seem to affect the prediction of the segmentation model.

To create the used synthetic datasets, CT acquired 3D anatomical models of the heart were used to extract the anatomical labels and create the input cases to the 3D GAN. The segmentation results and the echocardiography-looking aspect of the synthetic images pointed towards the generalization of this pipeline, as it can synthesize 3D echocardiography images, having as labels source different types of 3D models of the heart. The methodology to generate synthetic datasets can be generalized to other modalities, diseases, organs, as well as structures within the same organ (sub-regions of the heart, for example).

(Shin et al., 2018) and (Shorten and Khoshgoftaar, 2019) showed that GANs can be widely used to perform data augmentation of medical image datasets. The work from these authors, together with the presented results, encourage the main contributions of this work stating that GANs can be used to generate synthetic images with labels, working as a data augmentation strategy, and tackling the concern of scarcity of 3D echocardiography labeled datasets, especially if there are underrepresented data samples within the available real datasets.

### 4.3    3D Segmentation - Combined Datasets

Further results on the usage of synthetic datasets were explored and presented in Section III-D. Here, three datasets made of GAN generated and real 3D images were used to train more segmentation models and further evaluate the influence of the presence of real data in these datasets, as demonstrated in (Lustermans et al., 2022).

Fig. 6 a), b), and c) showed the anatomical segmentations of the LV, LA, and MYO predicted by the best trained fold of each model – $M_{Real}$, $M_{17Real10Augmented}$, and $M_{17Real20Augmented}$. From the qualitative analysis, the segmentations delineate well the anatomical structures in consideration throughout the whole volume. At the same time, and similarly to what was discussed on Section IV-B, the average Dice scores presented in Table 2 led to the conclusion that having a dataset of real images combined with synthetic ones leads to more accurate final segmentations.

From the obtained results is also possible to assess the influence of using combined datasets with different percentages of synthetic data. Table 2 and Table 3 show that adding synthetic data to the initial real dataset improves the 3D segmentation of real 3D echocardiography images. They also show that adding larger amounts of synthetic data does not improve the results to a large extent.

Fig. 6 d) showed the ground truth inference case segmentation performed by a cardiologist. From these ground truth segmentations available for all the cases in the test set, the Dice scores and the VS in Table 3 were calculated.

Given the 3D nature of the task and due to the Dice metric limitations, the VS was additionally calculated and used as comparison metric. In particular, $M_{17Real10Augmented}$ showed to perform better at segmenting when the Dice score was considered as performance metric. On the other hand, $M_{17Real20Augmented}$ performed better in terms of VS metric. These results showed that the models trained on the combined datasets, i.e., with real and synthetic images, provided more accurate segmentation outputs (the 3D volume), relatively to the model trained with only real data, $M_{Real}$. The results support the previous work done by (Lustermans et al., 2022), confirming that including synthetic data on datasets made of real data improves and helps the final outcome of the DL models.

Additionally, this result reinforces that the proposed pipeline, relying on a 3D GAN model, can be used as a data augmentation tool. This framework arises as a solution to the lack of publicly available medical labeled datasets.

### 4.4 Further Applications

The presented pipeline has the potential to be further explored. As the demand for medical images is increasing, the proposed approach can be extended to synthesize images from other imaging modalities other than Ultrasound, such as MR or CT. It can also generate images where other organs are represented or even fetuses (Roy et al., 2019).

Another extension of this work would be to use different types of 3D models from which ground truth anatomical labels could be extracted. Besides anatomical models, animated or biophysical models represent other options that can be considered. The usage of anatomical models of pathological hearts are another possible extension, in order to generate pathological 3D echocardiography scans. Depending on the type of 3D model being considered, different annotations can be extracted, increasing the amount of clinically relevant tasks where these synthetic datasets can be used.

The generated 3D echocardiography images illustrated a heart volume during one time step of the whole cardiac cycle (end-diastole). It would be of great interest to generate 3D images of the heart during other cardiac cycle events and even to generate a beating volume throughout time, as high temporal resolution is one of the main strengths of Ultrasound imaging. On the other hand, a limitation to the Ultrasound images generation is that different scanning probe combinations lead to the acquisition of images with different quality levels. This large variability makes the GAN learning process more complex.

In this work we explored, to an extent, the effects that architectural changes of the GAN model have on the final synthesized images. We used different architectures for the GAN generator but more 3D CNNs exist and are showing up every day. These can be used to train the generative models, since DL strategies are becoming extremely common to use as medical image synthesis and analysis tools. Once the images were synthesized, we used wavelet denoising and an in-house developed algorithm to fix the Ultrasound cone edges. However, there are other denoising transformations and cone reshaping algorithms that can be experimented to post-process the images.

We trained several DL models to perform 3D segmentation to show that synthesized images can be used as input to train DL models. Nevertheless, the pipeline is generic and could be applied to other DL tasks that automatically assign anatomical labels to images, e.g., structure/feature recognition or automatic structural measurements. Furthermore, the GAN-generated labeled datasets are not only useful as input to train DL models but also could be used to train researchers and clinicians on image analysis.

Finally, during this pipeline development, computational memory constraints were present, mainly due to the large size of 3D volumes, complicating the process of developing a framework adapted to these. Future work will include study strategies to overcome these limitations.

# 5 Conclusion

An automatic data augmentation pipeline to create 3D echocardiography images and corresponding anatomical labels using a 3D GAN model was proposed. DL models are becoming widely used in clinical workflows and large volumes of medical data is a fundamental requirement to develop such algorithms with high accuracy. Generating synthetic data that could be used for the purpose of training DL models is of utmost importance since this generative model can become a widely used tool to address the existent lack of publicly available data and increasing challenges with moving data due to privacy regulations. Furthermore, the proposed methodology not only generates synthetic 3D echocardiography images but also associates labels to these synthetic images, eliminating the need for experienced professionals to do so, and without adding potential bias in the labels.

The proposed GAN model shows a generalization component since it can generate synthetic echocardiography images using 3D anatomical models of the heart obtained for imaging modalities other than from Ultrasound.

The obtained results in this work indicate that synthetic datasets made up of GAN-generated 3D echocardiography images, and respective labels, are a good data augmentation resource to train and develop DL models that can be used to perform different medical tasks in the cardiac imaging domain, such as heart segmentation, where real patients' data is analyzed.

**Acknowledgements**

This work was supported by the European Union's Horizon 2020 research and innovation programme under the Marie – Skłodowska – Curie grant agreement No 860745.

**Appendix A**

See Table 4 and Table 5.

Table 4: Validation dice scores of each segmented structure (LV, LA, and MYO) for each trained model on combined datasets - M$_{Real}$, M$_{17Real10Augmented}$, and M$_{17Real20Augmented}$. The higher the score, the better the agreement between the model prediction and the ground truth segmentation. The best training fold of each model is highlighted.

| | \multicolumn{15}{c}{Models} | | | | | | | | | | | | | | |
|---|---|---|---|---|---|---|---|---|---|---|---|---|---|---|---|
| | $M_{Real}$ | | | | | $M_{17Real10Augmented}$ | | | | | $M_{17Real20Augmented}$ | | | | |
| | 1 | 2 | 3 | 4 | 5 | 1 | 2 | 3 | 4 | 5 | 1 | 2 | 3 | 4 | 5 |
| LV | 0.933 | 0.932 | **0.950** | 0.930 | 0.943 | 0.924 | 0.929 | 0.919 | **0.938** | 0.930 | 0.917 | 0.928 | **0.935** | 0.934 | 0.923 |
| LA | 0.837 | 0.869 | 0.873 | 0.837 | **0.896** | 0.830 | 0.841 | 0.820 | 0.808 | **0.853** | 0.831 | **0.841** | 0.838 | 0.829 | 0.793 |
| MYO | 0.710 | 0.699 | **0.766** | 0.697 | 0.750 | 0.745 | 0.745 | 0.779 | **0.815** | 0.735 | 0.715 | 0.771 | 0.766 | 0.780 | **0.785** |

Table 5: Validation dice scores of each segmented structure (LV, LA, and MYO) for each trained model on completely synthetic datasets – M$_{Synthetic}$, M$_{Wavelet}$, M$_{Cone}$, and M$_{WaveletCone}$. The higher the score, the better the agreement between the model prediction and the ground truth segmentation. The best training fold of each model is highlighted.

| | \multicolumn{20}{c}{Models} | | | | | | | | | | | | | | | | | | | |
|---|---|---|---|---|---|---|---|---|---|---|---|---|---|---|---|---|---|---|---|---|
| | $M_{Synthetic}$ | | | | | $M_{Wavelet}$ | | | | | $M_{Cone}$ | | | | | $M_{WaveletCone}$ | | | | |
| | 1 | 2 | 3 | 4 | 5 | 1 | 2 | 3 | 4 | 5 | 1 | 2 | 3 | 4 | 5 | 1 | 2 | 3 | 4 | 5 |
| LV | 0.924 | 0.930 | 0.924 | 0.918 | **0.934** | 0.927 | 0.930 | 0.923 | 0.919 | **0.934** | 0.926 | 0.930 | 0.921 | 0.918 | **0.933** | 0.928 | 0.928 | 0.914 | 0.914 | **0.935** |
| LA | **0.837** | 0.831 | 0.809 | 0.810 | 0.807 | 0.821 | **0.831** | 0.806 | 0.815 | 0.805 | 0.837 | **0.842** | 0.805 | 0.811 | 0.787 | **0.834** | 0.832 | 0.803 | 0.793 | 0.807 |
| MYO | **0.824** | 0.822 | 0.794 | 0.784 | 0.816 | **0.829** | 0.822 | 0.791 | 0.785 | 0.814 | **0.824** | 0.819 | 0.783 | 0.780 | 0.809 | **0.828** | 0.813 | 0.775 | 0.773 | 0.816 |